\documentclass{PoS}

\title{Lattice simulation of 2+1 flavors of overlap light quarks}

\ShortTitle{2+1 flavors of overlap light quarks}

\author{
  JLQCD collaboration:
  \speaker{S.~Hashimoto}$^{,a,b,}$\thanks{E-mail: shoji.hashimoto@kek.jp},
  S.~Aoki$^c$,
  H.~Fukaya$^d$,
  T.~Kaneko$^{a,b}$,
  H.~Matsufuru$^a$,
  J.~Noaki$^a$,
  T.~Onogi$^e$,
  N.~Yamada$^{a,b}$
  \vspace*{2mm}
  \\
  \llap{$^a$}
  High Energy Accelerator Research Organization (KEK),
  Tsukuba 305-0801, Japan
  \\
  \llap{$^b$}
  School of High Energy Accelerator Science,
  the Graduate University for Advanced Studies (Sokendai),
  Tsukuba 305-0801, Japan
  \\
  \llap{$^c$}
  Graduate School of Pure and Applied Sciences,
  University of Tsukuba, Tsukuba 305-8571, Japan
  \\
  \llap{$^d$}
  Theoretical Physics Laboratory, RIKEN, Wako 351-0198, Japan
  \\
  \llap{$^e$}
  Yukawa Institute for Theoretical Physics,
  Kyoto University, Kyoto 606-8502, Japan
}

\abstract{
  We report on the status of the dynamical overlap QCD simulation
  project by the JLQCD collaboration. 
  After completing two-flavor QCD simulation on a $16^3 \times 32$ lattice at
  lattice spacing $a\sim$ 0.12~fm, we started a series of runs with 2+1
  flavors. 
  In this report, we describe an outline of our algorithms, parameter choices,
  and some early physics results of this second phase of our project.  
}

\FullConference{The XXV International Symposium on Lattice Field Theory\\
		 July 30-4 August 2007\\
		 Regensburg, Germany}

\begin{document}

\section{Dynamical overlap fermion}
The JLQCD collaboration is carrying out a large scale lattice QCD simulation
using the overlap fermion formulation for sea quarks.
(An overview of the project has been given at this conference by Matsufuru 
\cite{Matsufuru_lat07}.)
The first phase of the project was a two-flavor QCD simulation on a 
$16^3\times 32$ lattice at a lattice spacing $a\simeq$ 0.11--0.12~fm.
The HMC simulations have been completed accumulating about 10,000 molecular
dynamics trajectories for six values of sea quark mass ranging $m_s/6$--$m_s$.
Preliminary reports of this project were already presented at Lattice 2006
\cite{Kaneko:2006pa,Matsufuru:2006xr,Hashimoto:2006rb,Yamada:2006fr}; 
at this conference we have presented physics results for 
pion masses and decay constants \cite{Noaki:2007es},
pion form factor \cite{Kaneko:2007nf},
kaon $B$ parameter \cite{Yamada:2007nh}, and 
topological susceptibility \cite{Chiu_lat07}.
We have also performed simulations in the $\epsilon$-regime by reducing the
sea quark mass down to 3~MeV.
This lattice has been used for the analysis of low-lying eigenvalues of the
overlap-Dirac operator \cite{Fukaya:2006xp,Fukaya:2007fb,Fukaya:2007yv} and
for a calculation of meson correlators in the $\epsilon$-regime
\cite{Fukaya_lat07}.
The second phase of the project is to include strange quark as dynamical
degrees of freedom: a 2+1-flavor QCD simulation with the overlap fermion.
We aim at producing dynamical lattices of size $16^3\times 48$ at around the
same lattice spacing $a\simeq$ 0.11--0.12~fm.

We use the Neuberger's overlap-Dirac operator
\cite{Neuberger:1997fp,Neuberger:1998wv} 
\begin{equation}
  \label{eq:ov}
  D(m) = \left(m_0+\frac{m}{2}\right)
  + \left(m_0-\frac{m}{2}\right)\gamma_5
  \mathrm{sgn}\left[H_W(-m_0)\right].
\end{equation}
The choice for the kernel operator is the standard Wilson fermion with a large
negative mass $m_0=1.6$.
For the gauge sector we use the Iwasaki gauge action together with extra
Wilson fermions and ghosts producing a factor
\begin{equation}
  \frac{\det\left[H_W(-m_0)^2\right]}{\det\left[H_W(-m_0)^2+\mu^2\right]}
\end{equation}
in the partition function such that the near-zero modes of
$H_W(-m_0)$ is naturally suppressed \cite{Fukaya:2006vs}.
This term is essential for the feasibility of dynamical overlap fermion
simulation, since it substantially reduces the cost of the approximation of
the sign function in (\ref{eq:ov}).
Although it prevents us from changing the topological charge during the
molecular dynamics evolutions, its systematic effect can be understood as a
finite size effect and can be estimated (and even corrected) once the
topological susceptibility is known \cite{Aoki:2007ka}.
The topological susceptibility is in fact calculable on the lattice with a
fixed topology as demonstrated in \cite{Chiu_lat07,Aoki:2007pw}.

\section{Algorithms}
For the calculation of the sign function in (\ref{eq:ov}) we use the rational
approximation 
\begin{equation}
  \label{eq:rational}
  \mathrm{sgn}\left[H_W\right] = H_W
  \left(p_0 + \sum_{l=1}^N \frac{p_l}{H_W^2+q_l}\right)
\end{equation}
with the Zolotarev's optimal coefficients $p_l$ and $q_l$.
This is applied after projecting out a few low-lying modes of $H_W$.
Typically, accuracy of order $10^{-(7-8)}$ is achieved with $N=10$.
The multiple inversions for $(H_W^2+q_l)^{-1}$ can be done at once using the
multi-shift conjugate gradient (CG).

The inversion of $D(m)$ is the most time-consuming part in the HMC
simulation. 
In the two-flavor runs, we mainly used the nested CG with relaxed 
residual for the inner CG \cite{Cundy:2004pza}.
In the 2+1-flavor runs, we use the five-dimensional solver as explained in the
following. 

By the Schur decomposition the overlap solver can be written in the form
(for $N=2$ for example) \cite{Narayanan:2000qx,Borici:2004pn,Edwards:2005an}
\begin{equation}
  \label{eq:5D}
  \left(
    \begin{array}[c]{cccc|c}
      H_W & -\sqrt{q_2} & & & 0 \\
      -\sqrt{q_2} & -H_W & & & \sqrt{p_2} \\
      & & H_W & -\sqrt{q_1} & 0 \\
      & & -\sqrt{q_1} & -H_W & \sqrt{p_1} \\
      \hline
      0 & \sqrt{p_2} & 0 & \sqrt{p_1} & R\gamma_5+p_0 H_W
    \end{array}
  \right)
  \left(
    \begin{array}[c]{c}
      \phi_{2+} \\
      \phi_{2-} \\
      \phi_{1+} \\
      \phi_{1-} \\
      \hline \psi_4
    \end{array}
  \right)
  =
  \left(
    \begin{array}[c]{c}
      0 \\
      0 \\
      0 \\
      0 \\
      \hline \chi_4
    \end{array}
  \right),
\end{equation}
where $R=(1+m)/(1-m)$.
By solving this equation we obtain a solution for $D(m)\phi_4=\chi_4$ with
$D(m)$ approximated by the rational function.
The matrix in (\ref{eq:5D}) can be viewed as a five-dimensional (5D) matrix.
An advantage of solving (\ref{eq:5D}) is that one can use the even-odd
preconditioning. 
Namely, rather than solving the 5D matrix $M$, we may solve a reduced matrix
$(1-M_{ee}^{-1}M_{eo}M_{oo}^{-1}M_{oe})\psi_e=\chi_e'$,
where even/odd blocks of $M$ are denoted by $M_{eo}$, $M_{ee}$, {\it etc}.
The inversion $M_{ee}^{-1}$ (or $M_{oo}^{-1}$) can be easily calculated
by the forward (or backward) substitution involving the 5D direction.

The low-mode projection can be implemented together with the 5D solver.
The lower-right corner is replaced by 
\begin{equation}
  R(1-P_H)\gamma_5(1-P_H)+p_0H_W+
  \left(m_0+\frac{m}{2}\right)\sum_{j=1}^{N_{ev}}
  \mathrm{sgn}(\lambda_j)v_j\otimes v_j^\dagger,
\end{equation}
where $P_H$ is a projector onto the subspace orthogonal to the $N_{ev}$
low-lying modes:
$P_H=1-\sum_{j=1}^{N_{ev}}\mathrm{sgn}(\lambda_j)v_j\otimes v_j^\dagger$.
Then, the inversion of $M_{ee(oo)}$ becomes non-trivial, but can be calculated
cheaply because the rank of the matrix is only $2(N_{ev}+1)$; 
the subspace is spanned by $x_e$, $\gamma_5x_e$, $v_{je}$, $\gamma_5 v_{je}$
($j=1, .., N_{ev}$).

\begin{figure}[tb]
  \centering
  \includegraphics[width=9cm,clip=true]{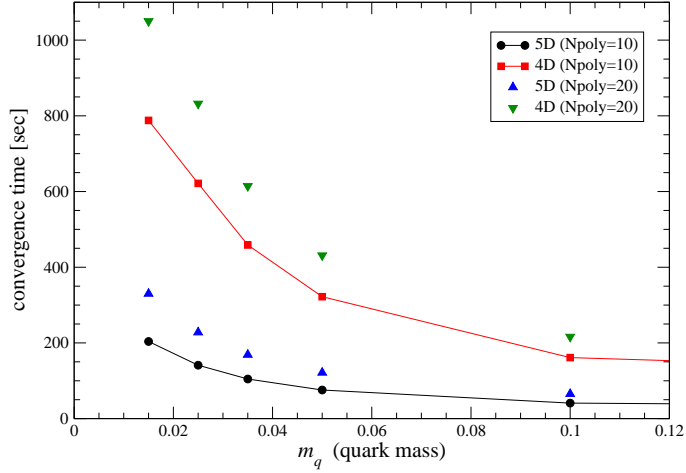}
  \caption{
    Comparison of solver performance.
    Data for $N=10$ is connected by lines: 
    4D (red squares) and 5D (black circles).
  }
  \label{fig:solver}
\end{figure}

We compare the performance of the 5D solver with the relaxed CG in 4D.
The elapsed time to solve the 5D equation is plotted in
Figure~\ref{fig:solver} as a function of quark mass $m$.
The lattice size is $16^3\times 48$ and the measurement is done on a half-rack
(512 nodes) of the BlueGene/L supercomputer (2.7~TFlops peak performance).
Data for $N=10$ is connected by lines for both 4D and 5D solvers.
Evidently, the 5D solver is faster by about a factor of 3--4.
Increasing the number of degree of the rational approximation requires more
computational cost for both 4D and 5D.
For the 5D case, the cost is naively expected to be proportional to $N$, but
the actual measurement shows slower increase, which indicates some overhead
due to the construction of low-mode projector {\it etc.}

\section{Odd number of flavors}
Introduction of the pseudo-fermions for dynamical quark flavors is the
starting point of HMC.
For the two-flavor case, this is straightforward by writing $\det D^2$ as
$\int[d\phi][d\phi^\dagger]\exp[-|H^{-1}\phi|^2]$,
where $H\equiv\gamma_5 D$.
The same trick applied for one flavor introduces $D^{-1/2}$ in the
pseudo-fermion action, which requires a method to calculate the inverse
square-root of the Dirac operator.
(For such algorithms, see \cite{Aoki:2001pt}, for example.)
For the overlap-Dirac operator this problem can be avoided as follows
\cite{Bode:1999dd,DeGrand:2006ws}.
Thanks to the exact chiral symmetry of the overlap fermion,
$H^2\equiv (\gamma_5 D)^2$ commutes with $\gamma_5$, and therefore can be
decomposed into positive and negative chirality subspaces:
\begin{equation}
  \label{eq:H2}
  H^2 = P_+ H^2 P_+ + P_- H^2 P_- \equiv Q_+ + Q_-,
\end{equation}
where $P_\pm = (1\pm\gamma_5)/2$.
Then, its determinant is factorized, 
$\det H^2 = \det Q_+ \cdot \det Q_-$.
Since $Q_+$ and $Q_-$ share the eigenvalues except for those of zero-modes,
$\det H^2 = (\det Q_+)^2 = (\det Q_-)^2$ up to the zero-mode contribution,
which is a trivial factor for the topology fixed simulations.
In order to simulate one flavor, one can just pick one chiral sector of $H^2$.

Thus, we introduce a pseudo-fermion field for the one-flavor piece as
$S_{PF1} = \sum_x \phi_\sigma^\dagger(x) Q_\sigma^{-1} \phi_\sigma(x)$, where
$\sigma$ can either be $+$ or $-$ representing the chiral sector.
At the beginning of each HMC trajectory, we refresh $\phi_\sigma(x)$ from a
gaussian distribution $\xi(x)$ as $\phi_\sigma(x)=Q_\sigma^{-1/2}\xi(x)$.
This step requires a calculation of the square-root of $Q_\sigma$, which is
done using the rational approximation.
Calculation of the molecular-dynamics force is straightforward: one can simply
project onto the chiral sector $\sigma$ in the calculation of the force from
$H^2$. 

\section{Runs}
The 2+1-flavor runs are done at $\beta=2.30$, which is the same value as our
main two-flavor runs.
The unit trajectory length $\tau$ is set to 1.0, twice longer than the
two-flavor runs.
Our choice of the sea quark mass parameters are summarized in
Table~\ref{tab:para}. 
The up and down quark mass $m_{ud}$ ranges from $m_s$ down to $\sim m_s/6$ as
in our two-flavor runs.
For the strange quark mass we take two values aiming at interpolating to the
physical strange quark mass.

\begin{table}[tb]
  \centering
  \begin{tabular}{|c|c|c|}
    \hline
    $m_{ud}$ & $m_s$ = 0.080 & $m_s$ = 0.100 \\
    \hline\hline 0.015 & $\surd$ & $\surd$ \\
    \hline       0.025 & $\surd$ & $\surd$ \\
    \hline       0.035 & $\surd$ & $\surd$ \\
    \hline       0.050 & $\surd$ & $\surd$ \\
    \hline       0.080 & $\surd$ &   \\
    \hline       0.100 &   & $\surd$ \\
    \hline
  \end{tabular}
  \caption{Sea quark mass parameters}
  \label{tab:para}
\end{table}

At the time of the lattice conference, the runs proceeded to 500--1,000 HMC
trajectories depending on the mass parameter.
One trajectory takes about 1--2 hours on one rack (1,024 nodes) of BlueGene/L
(5.7~TFlops peak performance).
The acceptance rate is kept around 80--90\% for each run.

\begin{figure}[tb]
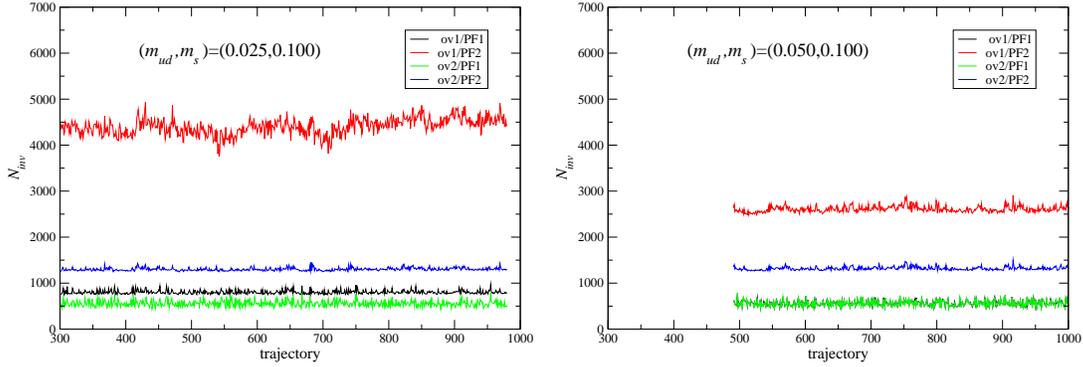

  \centering
  \includegraphics[width=7cm,clip=true]{conv_0.025_0.100.eps}
  \ \ 
  \includegraphics[width=7cm,clip=true]{conv_0.050_0.100.eps}
  \caption{
    Molecular dynamics time evolution of the number of CG iterations in the
    calculation of the HMC Hamiltonian.
    Data at $m_{ud}=0.025$ (left) and 0.050 (right) with $m_s=0.100$.
    In the plot ``ov1'' denotes up and down quarks, while ``ov2'' corresponds
    to strange.
    ``PF2'' stands for the inversion with the original sea quark mass, and
    ``PF1'' is for the preconditioner, whose mass is chosen to be
    0.4 for $m_q\ge 0.035$ or 0.2 for $m_q\le 0.025$.
  }
  \label{fig:ninv}
\end{figure}

Figure~\ref{fig:ninv} shows the number of the (two) 5D CG iteration in the
calculation of the HMC Hamiltonian.
As expected the calculation for the two-flavor piece is dominating the
calculation.

\begin{figure}[tb]
  \centering
  \includegraphics*[width=9cm,angle=0,clip=true]{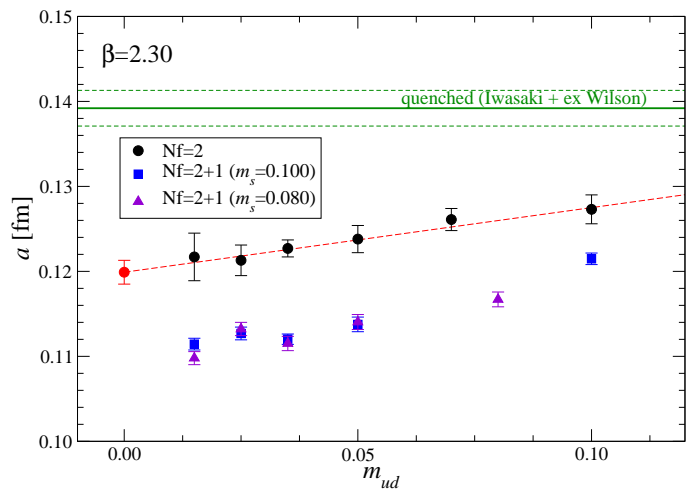}
  \caption{
    Lattice spacing as a function of sea quark mass.
    At $\beta$ = 2.30, two-flavor data (black circles) are plotted together
    with a line of chiral extrapolation.
    2+1-flavor data are plotted for both $m_s$ = 0.100 (blue squares) and 0.080
    (blue triangles).
    A quenched result at the same $\beta$ value is shown by a red band.
  }
  \label{fig:a}
\end{figure}

Measurements of physical quantities are done at every 5 trajectories, so far
only for the $m_s=0.100$ lattices.
In order to use in the low-mode preconditioning and low-mode averaging, we
are calculating 80 pairs of low-lying eigenmodes of the overlap-Dirac
operator. 
The lattice spacing as determined through the Sommer scale $r_0$ (= 0.49~fm)
is plotted in Figure~\ref{fig:a} for both 2-  and 2+1-flavor lattices.
At the same $\beta$ value (= 2.30) the lattice spacing decreases as more
dynamical flavors are included.

\begin{figure}[tb]
  \centering
  \includegraphics[width=7cm,clip=true]{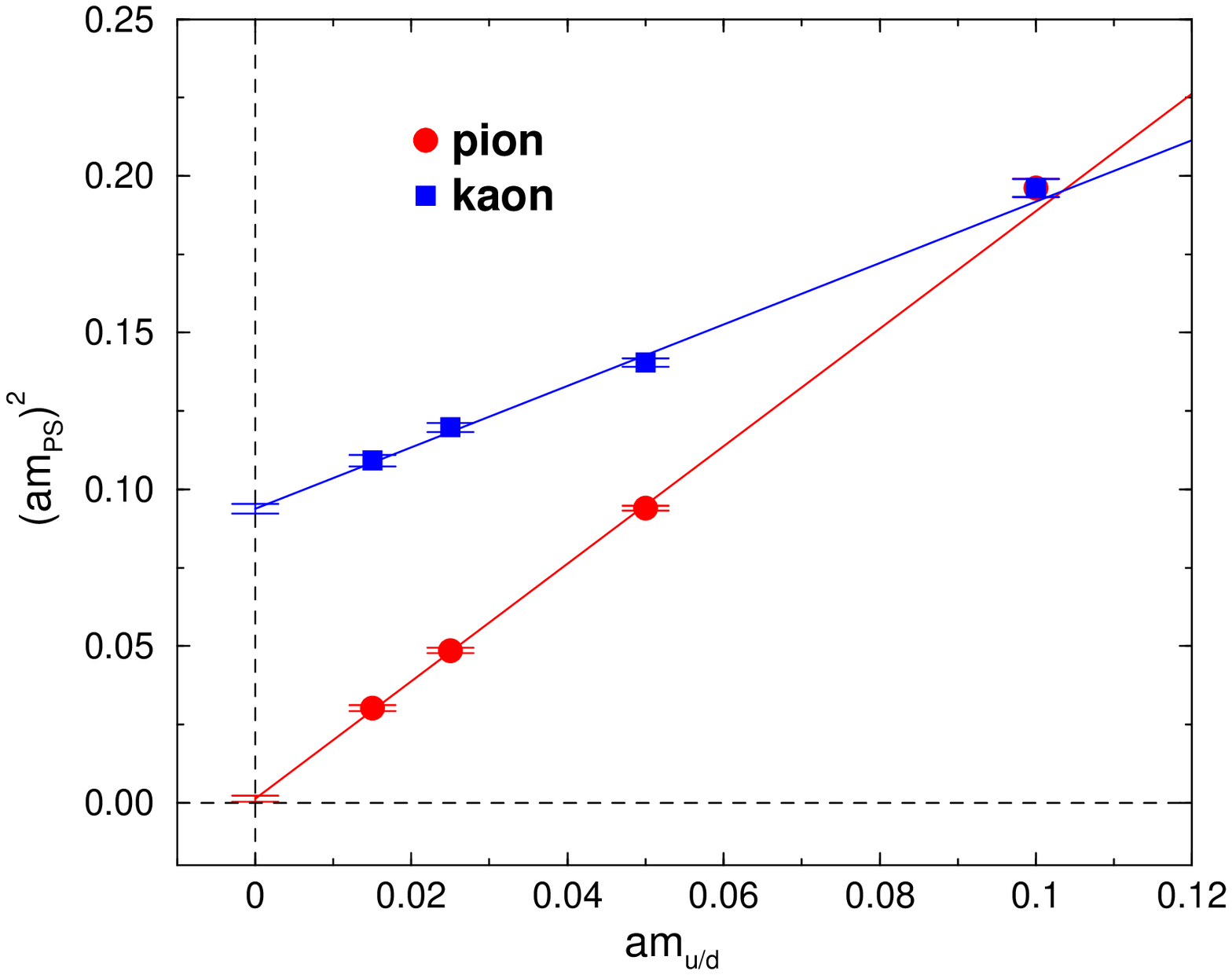}
  \ \
  \includegraphics[width=7cm,clip=true]{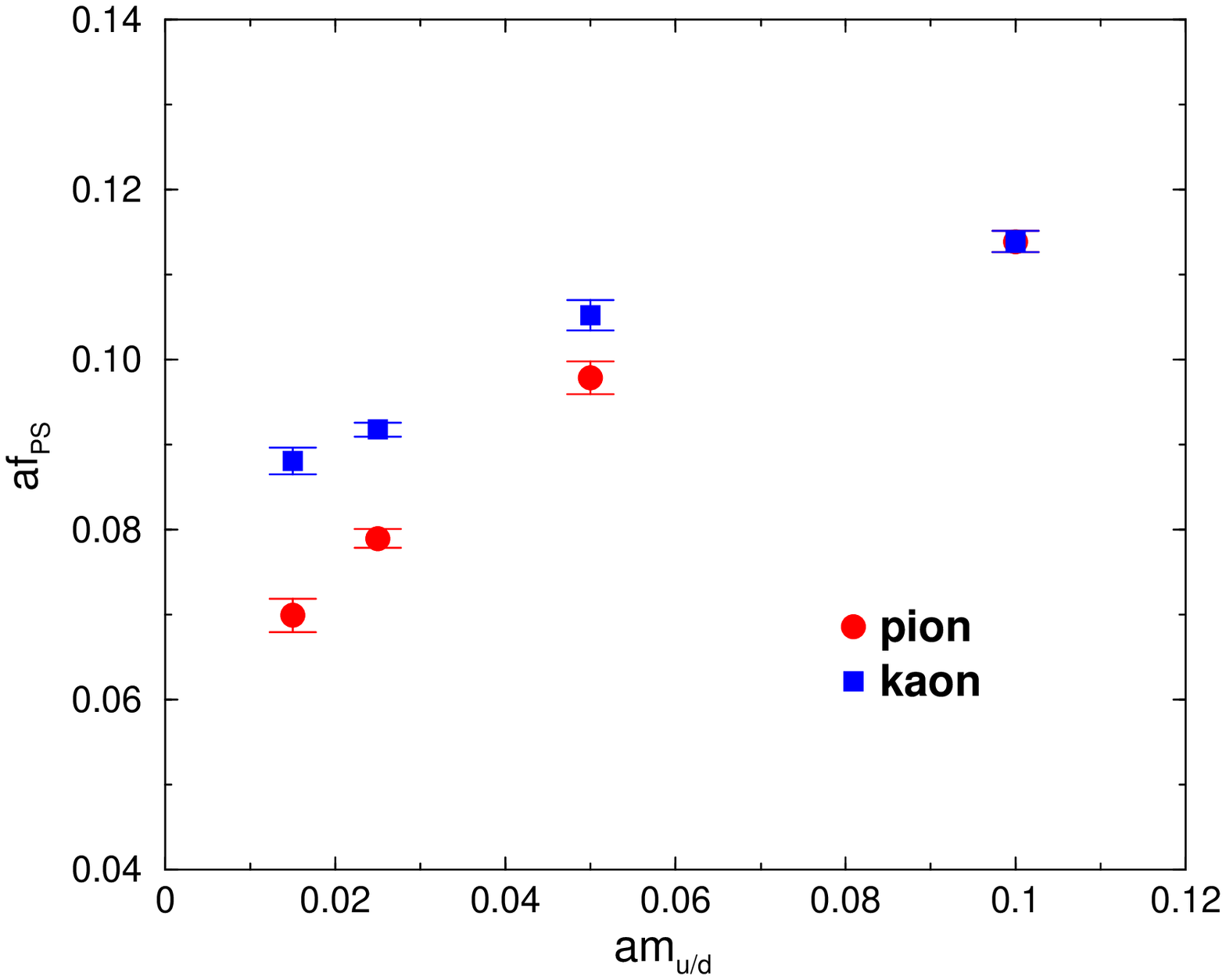}
  \caption{
    Preliminary results for pion and kaon mass squared (left)
    and their decay constants (right) as a function of sea quark mass.
  }
  \label{fig:results}
\end{figure}

Preliminary results for pion and kaon mass squared and decay constant are
shown in Figure~\ref{fig:results}. 
Data at $m_s=0.100$ are plotted as a function of sea quark mass.
Although the statistics is still low ($<$ 1,000 trajectories for each sea
quark mass), reasonably precise data are obtained using the low mode averaging
technique. 
Detailed analysis with the chiral extrapolation is yet to be done after
accumulating more statistics.

\vspace*{1cm}
Numerical simulations are performed on Hitachi SR11000 and IBM System Blue
Gene Solution at High Energy Accelerator Research Organization (KEK) under a
support of its Large Scale Simulation Program (No.~07-16).
This work is supported in part by the Grant-in-Aid of the Ministry of
Education, Culture, Sports, Science and Technology 
(No.~17740171, 18034011, 18340075, 18740167, 18840045, 19540286 and 19740160).

\end{document}